\documentclass[12pt,letter,hyper]{JHEP3}
\usepackage{amsmath}
\usepackage{amscd}
\usepackage{amsfonts}
\usepackage{amssymb}
\usepackage{amsthm}

\usepackage{cite}
\usepackage{epsfig}

\newcommand {\CalF} {\mathcal F}

\newcommand {\CalJ} {\mathcal J}

\newcommand {\CalN} {\mathcal N}

\newcommand {\BR}   {\mathbb R}
\newcommand {\BZ}   {\mathbb Z}
\newcommand {\BC}   {\mathbb C}

\newcommand {\p} {\partial}

\DeclareMathOperator{\End}{End}

\DeclareMathOperator{\I} {Im} \DeclareMathOperator{\R} {Re}
\DeclareMathOperator{\Pf} {Pf}

\title{Black hole entropy and  topological strings on generalized CY manifolds }

\author{Vasily Pestun\thanks{On leave of absence from ITEP, 117259, Moscow, Russia}\\
Physics Department, Princeton University, Princeton NJ, 08544 }

\preprint{ ITEP-TH-110/05 \\PUPT-2188
\\ }

\abstract{
  The H. Ooguri, A. Strominger and C. Vafa conjecture $Z_{BH}=|Z_{top}|^2$ is
extended for the topological strings on generalized CY manifolds.
It is argued that the classical black hole entropy
is given by the \emph{generalized Hitchin functional}, which defines by critical points
a generalized complex structure on $X$. This geometry differs
from an ordinary geometry if $b_1(X) \neq 0$.
In a critical point
the generalized Hitchin functional equals to Legendre transform
of the free energy of generalized topological string. The examples of $T^6$ and $T^2 \times K3$
are considered in details.

}

\begin{document}
\section{Introduction}

In~\cite{Ooguri:2004zv} H. Ooguri, A. Strominger and C. Vafa (OSV) suggested
a relation between the black hole entropy and the topological string
partition function symbolically written as $Z_{BH}=|Z_{top}|^2$.
In~\cite{Nekrasov:2004vv} N. Nekrasov and in~\cite{Dijkgraaf:2004te} R. Dijkgraaf et. al
explained that at the classical level the black hole entropy and
the topological strings partition function
 are related to a certain Hitchin functional~\cite{Hitchin} for real three-forms,
which defines by critical points a CY structure on a real six-dimensional
manifold~$X$, so $Z_{hit} = Z_{BH} =|Z_{top}|^2$.
The important relation between Hitchin functional~\cite{Hitchin}
and the quantization of the topological $B$-model~\cite{Bershadsky:1993cx}
was shown in~\cite{Gerasimov:2004yx} by A.~Gerasimov and S.~Shatashvili.
Moreover, in~\cite{Gerasimov:2004yx}
was also suggested to use the generalized Hitchin functional~\cite{Hitchin:2}, whose
degrees of freedom are extended by one-forms and five-forms on $X$.
The necessity to turn on forms of all ranks was proposed in~\cite{Nekrasov:2004vv}
in the perspective of a certain seven-dimensional topological theory.
See~\cite{Pestun:2005gs} for topological strings in generalized complex space~\cite{Hitchin:2,GCS}.

The whole construction is about compactifications with (at least) $\CalN=2$ supersymmetry.
The usual ones are compactifications on Calabi-Yau manifolds. Then one
has the $A$-model and the $B$-model~\cite{Witten:1991zz},
parameterized by symplectic or complex structures.  They are each
examples of generalized complex
structures~\cite{Hitchin:2,GCS} and each can be described by a suitable
Hitchin functional~\cite{Hitchin,Hitchin:2}.
The black hole entropy in the
supergravity approximation is equal to
the Hitchin functional~\cite{Ooguri:2004zv,Nekrasov:2004vv,Dijkgraaf:2004te}.

At the one-loop level, however, to reproduce the first quantum
correction to the Hitchin functional,
one needs to use the generalized Hitchin functional, as was shown in~\cite{Pestun:2005rp}.
In other words, for Calabi-Yau compactification, at tree level, only the modes of
the ordinary Hitchin functional
are turned on, but at one-loop level, to get the right quantum
correction, one needs to allow the extra
fields of the generalized Hitchin functional to run around in the loop.

The present paper is devoted to answering the following question:  Is it
possible to find a situation in which
it is necessary to use the generalized Hitchin functional at tree level in
order to reproduce the supergravity
approximation to the black hole entropy?    This will happen if the
extra fields of the generalized Hitchin
functional have expectation values at tree level.

In other  words, it will happen in the case of a compactification with at
least $\CalN=2$ supersymmetry that
cannot be described as compactification on a complex manifold ($B$-model)
or a symplectic manifold ($A$-model)
-- compactification that requires the language of a generalized complex
structure.
Concretely, this will happen in compactifications of $\CalN\geq 2$
supersymmetry on a manifold X with $b_1(X)$ nonzero.

There is a consistent supergravity analysis of $\CalN=2$ supersymmetric
compactifications with $b_1(X)$ nonzero
and using generalized complex geometry~\cite{Witt:2005oct,Grana:2005jc,Grana:2005ny,Grana:2004bg,Chuang:2005qd,Lindstrom:2004iw,Fidanza:2003zi}.  In this paper, we will show
that in this situation, the generalized Hitchin functional reproduces
the supergravity approximation to the black
hole entropy, generalizing the results of OSV for the Kahler case.

However, actual examples of the framework of~\cite{Grana:2005jc,Witt:2005oct,Grana:2005ny,Grana:2004bg,Chuang:2005qd,Lindstrom:2004iw,Fidanza:2003zi} are apparently not yet known.
To give concrete examples of our calculations, therefore, we will
consider the examples of
$X=T^6$ and $X=T^2 \times K3$, which certainly do have $b_1(X)$ nonzero.  They have
more than $\CalN=2$ supersymmetry,
so it may be the case that at the loop level, they cannot be
described by the generalized Hitchin
functional (but require some further extension of it with additional
fields related to the higher supersymmetry).
However, at tree level compactification on $T^6$ or $T^2 \times K3$ has a
consistent truncation with $\CalN=2$ supersymmetry
that includes deformations best described by generalized complex
geometry.  In this paper, we will
show that in this subspace of the $T^6$ and $T^2 \times K3$ moduli space, the
black hole entropy at tree level
is described by the generalized Hitchin functional.    We hope that in
the future examples will be found
illustrating the ideas of the framework of~\cite{Grana:2005jc,Witt:2005oct,Grana:2005ny,Grana:2004bg,Chuang:2005qd,Lindstrom:2004iw,Fidanza:2003zi,Gurrieri:2002wz} with $\CalN=2$ supersymmetry.





For generalized complex space, the necessary formalism
of topological strings -- topological $\CalJ$-model -- is presented
in~\cite{Pestun:2005gs}, see also~\cite{MR1609624,MR1919435,MR1432574,Kapustin:2004gv,Kapustin:2003sg,Li:2005tz,Zabzine:2005qf,Lindstrom:2004hi,Lindstrom:2004iw,Zucchini:2005cq,Zucchini:2005rh,Zucchini:2004ta}.
The generalized Hitchin functional in~\cite{Hitchin:2},
and the compactifications of type II string theory on generalized CY
manifolds are studied in~\cite{Grana:2005ny,Grana:2005jc,Grana:2004bg,Chuang:2005qd,Lindstrom:2004iw,Fidanza:2003zi}.
For recent developments on black hole
entropy see~\cite{Dabholkar:2004yr,Dabholkar:2005dt,Dabholkar:2005by,Parvizi:2005aa,Sen:2005ch,Sen:2005wa,Sen:2005iz,Verlinde:2004ck,Strominger:1996kf,Strominger:1996sh,Horowitz:1996fn, Maldacena:1996gb,Breckenridge:1996sn,Kaplan:1996ev,Maldacena:1997de,Maldacena:1999bp,Ferrara:1995ih} and on generalized
complex structures in string theory
see~\cite{Lindstrom:2004hi,Zucchini:2005rh,Zucchini:2005cq,Lindstrom:2004eh,
Zucchini:2004ta,Kapustin:2004gv,Chiantese:2004pf,Zabzine:2004dp,Gates:1984nk}.

In sec.~\ref{sec:gen} we briefly review the standard logic,
in sec.~\ref{sec:main} we show that it is easily  generalized. In sec.~\ref{sec:ex}
we illustrate an emergence of the generalized Hitchin
 functional for $T^6$ and $T^2 \times K3$ compactifications.
The sec.~\ref{sec:con} concludes the note.

\section{A review}\label{sec:gen}

The relation between $Z_{BH}$ and $Z_{top}$ comes from considering
a compactification of the physical type II string on a  Calabi-Yau threefold $X$.
See~\cite{Mohaupt:2000mj} for a comprehensive review of the subject and the complete list of references.

The resulting low energy effective theory is $\CalN=2$
four-dimensional supergravity. It contains the $\CalN=2$
gravitational multiplet, the universal hypermultiplet and a number
of vector and hyper multiplets depending on the geometry of $X$.
For type IIA string $h^{1,1}$ vector multiplets correspond to
the complexified Kahler moduli of $X$, and $h^{2,1}$ hyper
multiplets correspond to the complex moduli of $X$. For type IIB
string the structure is reversed. The low energy effective action
for vector multiplets is fully specified by a single
holomorphic\footnote{In homogeneous special coordinates, $\CalF$ is
a homogeneous function of weight 2.} function, the prepotential
$\CalF(X^I)$. The $X^I$ are scalar components of the vector
multiplets, they describe  moduli of the CY manifold $X$. The
prepotential $\CalF(X)$ defines a structure of the special Kahler
geometry on the corresponding moduli space.

On the one hand, the physical string amplitudes on $X$, which compute $\CalF(X)$, can be formulated
in the language of topological strings~\cite{Antoniadis:1993ze,Bershadsky:1993cx}. Namely,
$\CalF(X)$ is just the classical free energy of the topological string. The higher
genus amplitudes give the terms
\begin{align}
    \label{I_g} I_g = \int d^4 \theta \, W^{2g}F_g(X^I),
\end{align}
where $X^k$ are the $\CalN=2$ chiral superfields constructed
from the vector multiplets, and $W$ is the $\CalN=2$ chiral superfield for the Weyl
multiplet $W^{ij}_{\mu\nu} = T^{ij}_{\mu\nu} -
R_{\mu\nu\l\rho}\theta^i \sigma_{\l\rho} \theta^j + \cdots $ (with
$T$ being the graviphoton field, so the expansion in components
of~\eqref{I_g} gives terms $R^2T^{2g-2}$).

On the other hand, the four-dimensional $\CalN=2$ supergravity
admits BPS black hole solutions~\cite{Ferrara:1995ih,Strominger:1996kf,Sabra:1997dh}.
These BPS black hole solutions are generalizations of the extremal
Reissner-Nordstrom  black holes in Einstein-Maxwell theory with $M = |Q|$.
The Reissner-Nordstrom black hole has the metric
\begin{align}
 ds^2 = -dt^2 (1 - 2M/r + Q^2/r^2) + dr^2 (1-2M/r + Q^2/r^2)^{-1} + r^2
 d\Omega_2^2.
\end{align}
There is a bound $|Q| \leq M$. When the bound $M=Q$ is
reached, the solution becomes BPS solution. The BPS solution preserves $N=1$ supersymmetry.
Moreover, the near horizon geometry of such a solution is given by the Bertotti-Robinson
metric $AdS_2 \times S^2$. In coordinates, where horizon is located at $r=0$, the metric
is given by
\begin{align}
 ds^2 = -\frac {r^2} {Q^2} dt^2 + \frac{Q^2} {r^2} (dr^2 + r^2
 d\Omega^2).
\end{align}
The radius of this black hole is $r_0 = M = |Q|=|Z|$, where we also introduced
the central charge of $\CalN=2$ algebra for such a BPS object.
The Bekenstein-Hawking-Wald
 entropy~\cite{Bardeen:1973gs,Bekenstein:1974ax,Bekenstein:1973ur,Wald:1993nt} is given by
the familiar formula
\begin{align}
  S = \frac {1} {4} Area = \pi r_{0}^2 = \pi |Q|^2.
\end{align}

In the full $\CalN=2$ supergravity we turn on the abelian vector multiplets.
Each one has a complex scalar $X^I$ and  magnetic and electric fields $F^+_{\mu\nu}{}^I,
G^+_{\mu\nu}{}^I$. In the language of special Kahler geometry, it is convenient
to organize the fields into pairs $(X^I, F_I := \p_I \CalF)$ and $(F^+,G^+)$
 that transform linearly under $Sp(2n+2,\BR)$ duality group,
actually broken to $Sp(2n+2,\BZ)$:
\begin{align}
 \left(\begin{array}[2]{c}
   X^I \\
   F_I
 \end{array}\right) \quad \quad
 \left(
\begin{array}[2]{c}
   F^+_{\mu\nu}{}^I \\
   G^+_{\mu\nu}{}_I
 \end{array}\right).
\end{align}
The black hole can carry magnetic and electric charges $(p^I, q_I)$, and the solution
is given by the usual $\frac 1 {r^2}$ law in the metric $ds^2 = -e^{2g(r)} dt^2 + e^{2f(r)}[dr^2+r^2 d\Omega_2^2]$
\begin{align}
 i F_{23}^I = i \frac {e^{-2f(r)}} {r^2} p^I, \quad i G_{23I} = i \frac {e^{-2f(r)}} {r^2} q_I.
\end{align}
It is convenient to introduce the central charge field $Z = e^{K/2} (p^I F_I - q_I X^I)$,
where the Kahler potential $ e^{-K} = 2 \I (X^I \bar F_I)$.
The supersymmetry condition gives the solution for the
metric in terms of $Z$, so we have $e^{2g(r)} = e^{-2f(r)} = e^{-K}\frac{r^2}{|Z|^2}$.
From the Wald formula for the entropy one obtains again
\begin{align}
\label{eq:BH}
 S = \pi |Z|^2.
\end{align}
Since $Z$ is expressed in terms of the scalars $(X^I,F_I)$ and
the charges $(p^I,q_I)$ we still need to find $(X^I,F_I)$
in terms of $(p^I,q_I)$ and then plug into~\eqref{eq:BH}. The relation is given
by the so called attractor equations\footnote{See~\cite{Moore:1998pn} for studies of their relation to number theory.}~\cite{Ferrara:1996um,Ferrara:1996dd,Ferrara:1995ih,Strominger:1996kf,LopesCardoso:1998wt,LopesCardoso:1999cv}
\begin{align}
 \bar Z \left(\begin{array}[2]{c}
   X^I \\
   F_I
 \end{array}\right) - Z
 \left(
\begin{array}[2]{c}
   \bar X^I  \\
   \bar F_J
 \end{array}\right)
= i e^{-K/2}
 \left( \begin{array}[2]{c}
   p^I \\
   q_J
 \end{array}\right).
\end{align}
So we have the formula for the entropy $S(p,q)$
\begin{align}
\label{eq:Spq}
  S(p^I,q_I) = \pi \frac{|pF - qX|^2}{ 2 \I (X \bar F)}.
\end{align}
where $\R (CX^I) = p^I$, $\R (CF_I) = q_I$ (we suppress index $I$ in contractions like $X^I \bar F_I$).
The formula~\eqref{eq:Spq} is invariant under a homogeneous complex dilatation,
so we can put $C=1$. The attractor equations $\R (CX^I) = p^I, \R (CF_I) = q_I$ can be also
obtained minimizing~\eqref{eq:Spq} by $X^I$ for the fixed charges $(p^I,q_I)$
with $F_I  = \p_I \CalF$.

Let us decompose $X^I$ into the imaginary
and the real part\footnote{In the following we often omit the index $I$,
assuming the natural contraction in products.} $X' + i X'', F = F' + iF''$.
Then we  plug $p = X', q = F'$ and compute $S(p,q) = S(X',F')$
\begin{align}
 S_{BH}(X',F') = \pi \frac{ |X'(F'+iF'') - F'(X'+iX'')|^2  }{2( X'' F' - X' F'')} = \frac {\pi}{2} (X'' F' -
 X'F'').
\end{align}

Now compare the function $S_{Hit}$, whose rationale  will become clear in a moment,
$S_{Hit}(X',F') = \frac {1} {\pi} S_{BH}(X',F')$
with the imaginary part or the prepotential $\CalF= \frac 1 2 X^I F_I = \CalF' + i \CalF''$
\begin{align}
S_{Hit}(X',F') = \frac 1 2 (X'' F' - X' F'') \\
\CalF''(X',X'') = \frac 1 2 ( X''F' + X' F'').
\end{align}
We see that
\begin{align}
 \frac 1 {\pi} S_{BH}(X',F') = S_{Hit}(X',F') = X''F' - \CalF''(X',X'').
\end{align}
Moreover, since $\CalF = \frac 1 2 X^I F_I$ and $\CalF$ is holomorphic we have the relation on the derivative
\begin{align}
  F'_I = \frac{ \p \CalF''} {\p X^I{}''}.
\end{align}
Therefore $\frac 1 {\pi} S_{BH} (X',F') = S_{Hit}(X',F')$ is
the Legendre transform of the imaginary part
of the topological string free energy $ \CalF''(X',X'')$ in the imaginary part  $X''\equiv \I
X$~\cite{Ooguri:2004zv,Dijkgraaf:2004te,Nekrasov:2004vv}
\begin{align}
\label{eq:HitLeg}
\boxed{
  S_{Hit} (X',F') = \mathrm{Legendre} [\CalF''(X',X''),  F' =\p_{X''} \CalF'' ]}
\end{align}

\section{A generalization of the OSV conjecture}
\label{sec:main}
Before going to generalization, let us recall the meaning of the Hitchin functional $S_{Hit}$
in the formulas above. Let $\Omega$ be the holomorphic $(3,0)$ form on the CY manifold $X$.
As usual we have $X^I = \int_{A_I} \Omega, F_I = \int_{B^I} \Omega$ for some canonical basis
of cycles $A_I, B^I$. The Hitchin functional in its critical point is the
integral of the volume form defined by $\Omega$
\begin{align}
  S_{Hit} (X', F') = - \frac {i} {4} \int \Omega \wedge \bar \Omega = \frac {1} {4 i} (X \bar F - \bar XF)=
\frac 1 2 \I X\bar F.
\end{align}
The reason why we write $S_{Hit}$ as a function of real part of periods $X',F'$ is that
it is actually a function of them by the construction~\cite{Hitchin}
\begin{align}
 S_{Hit}[\rho] = \frac {1} {4i} \int (\rho + i\hat \rho ) \wedge (\rho -i \hat \rho) =
-\frac {1} {2} \int ( \rho \wedge \hat \rho) = \int \sqrt{I_4(\rho)} = \int vol.
\end{align}
Here $\rho$ is a stable real three-form, and $\hat \rho$ is a
certain non-linear function of $\rho$, such that $\rho+i \hat
\rho$ is the decomposable  almost holomorphic $(3,0)$ form with
respect to the complex structure also defined by $\rho$. The
integrability of the complex structure can be cast in the form $d
(\rho + i \hat \rho) =0$. The field theory is defined by
restricting $\rho$ to some cohomology class in $H^3(X,\BR)$, so $d\rho = 0$.
In a critical point of $S_{Hit}[\rho]$ we have $d \hat \rho =
0$, and the complex structure is integrable.
We see that at the classical level the relation~\eqref{eq:HitLeg} holds: the Hitchin
functional, proportional to the black hole entropy, is the
Legendre transform of the imaginary part of the holomorphic
prepotential~$\CalF$~\cite{Ooguri:2004zv,Dijkgraaf:2004te,Nekrasov:2004vv}.
The relation between Hitchin functional and topological
string was also studied
classically in~\cite{Gerasimov:2004yx} and at the one-loop in~\cite{Pestun:2005rp}.
For micro/macroscopical tests of the OSV conjecture see~\cite{Dabholkar:2004yr,Dabholkar:2005dt,Dabholkar:2005by,Verlinde:2004ck,Strominger:1996kf,Strominger:1996sh,Horowitz:1996fn, Maldacena:1996gb,Breckenridge:1996sn,Kaplan:1996ev,Maldacena:1997de,Maldacena:1999bp,Ferrara:1995ih}.

In the case of generalized complex structures the whole construction works exactly in the same
way. For a generalized complex structure, an analogue
of the holomorphic $(3,0)$-form will be a mixed differential form
in complex $H^{odd} = H^1 + H^3 + H^5$
or $H^{even} = H^0 + H^2 + H^4 + H^6$,
which is at the same time a pure spinor $\Omega=\rho + i \hat \rho$ of $SO(6,6)$~\cite{Hitchin:2,GCS}.
The off-shell generalized Hitchin functional is defined by the
real part $\rho$ of the pure spinor~$\Omega$
\begin{align}
\label{eq:HitGen}
S_{Hit}^G = -\frac {1} {2} \int (\rho,\hat\rho) = \int
\sqrt{I_4(\rho)},
\end{align}
where $\hat \rho$ is a certain nonlinear function of $\rho$, and $(,)$ is an appropriate bilinear
form on the space of mixed differential forms~\cite{Hitchin:2}.
A mixed differential form $\rho$ in $\Omega^1+\Omega^3+\Omega^5$
or $\Omega^0+\Omega^2+\Omega^4+\Omega^6$, according to its chirality,
 transforms as a spinor of $SO(6,6)$, and $I_4(\rho)$
is the singlet in the tensor product of four $SO(6,6)$ spinors.

The moduli space of ordinary CY structures
locally is
\[(H^{3,0}\oplus H^{2,1})(X,\BC),\]
or $H^3(X,\BR)$ by Hitchin construction.
The moduli space of
generalized CY structures locally near the point of an ordinary
complex structure is
\[(H^{1,0} \oplus H^{2,1} \oplus H^{3,2} \oplus H^{3,0})(X,\BC),\]
or $H^1(X,\BR) \oplus H^3(X,\BR) \oplus H^5(X,\BR)$ by Hitchin
construction.

The $even/odd$ cases of generalized complex structure in six real dimensions
correspond to the type A/B
strings. They are distinguished by the chirality of the canonical pure spinor $\Omega$
that defines the corresponding generalized complex structure. In real six dimensions
a usual complex structure is of odd type, and a usual symplectic structure is of even type.

In \cite{Hitchin:2} Hitchin shows that the moduli
space of generalized complex structures has \emph{a special Kahler geometry}.
Since $\CalN=2$ supergravity is fully defined by an appropriate special Kahler structure
on the target manifold for the scalar fields from vector multiplets,
all $\CalN=2$ computations for the black hole entropy can be done in
the generalized complex case, as long as one includes the extra multiplets.

The outcome of $\CalN=2$ supergravity is the
formula~\eqref{eq:HitLeg},
which tells us that $S_{BH}$ is the Legendre transform of $\I \CalF$. Here $\CalF$ is the
prepotential of the special geometry of the moduli space of generalized complex structures.
It can be defined in a similar way.  We pick up a basis of $A_I,B^I$ cycles in
$H_{odd}=H_1 \oplus H_3 \oplus H_5$ or in $H_{even}=H_0 \oplus H_2 \oplus H_4 \oplus H_6$,
which is canonical with respect to the \emph{sign twisted} wedge product\footnote{
This sign twisted wedge
product for two forms $\alpha, \beta$ is defined as  $(\alpha, \beta) = \alpha \wedge \beta$
for $\deg \beta = 4k+0,1$ and $(\alpha,\beta) = -\alpha \wedge \beta$ for $\deg \beta = 4k+2,3$ \cite{Hitchin:2}.}
that agrees with the bilinear form on spinors of $Spin(TX,TX)$~\cite{Hitchin:2}.
Then
\begin{align}
  X^I = \int_{A_I} \Omega,\quad   F_I \int_{B^I} \Omega,
\end{align}
where now $A_I,B^I$ runs over all degrees in $H_{odd}$ of $H_{even}$.
For example, let us consider an ordinary symplectic structure $\omega$ as a generalized complex structure.
Then $\Omega = e^{i\omega}$, or
\[\Omega = 1 + i\omega -\frac 1 2 \omega^2 - \frac 1 6 i \omega^3.\]
We have the zero-cycle and a number of two-cycles of $A$ type, and a number four-cycles and the
six-cycle of $B$ type.
The sign twisted wedge product is antisymmetric and defines
a symplectic structure on~$H^{even}(X)$. Then we recover the standard formulas
\begin{align}
X^0 =1 \quad \quad  F_0 = \int_X -i \frac 1 6 \omega^3\\
X^I =  \int_{A_I} i \omega \quad  \quad F_I = \int_{B^I} - \frac 1 2 \omega^2\\
\quad  \CalF = \left(\frac{-i} {4}+ \frac {i} {12} \right) \int_X  \omega^3 = - \frac {i} {6} \int_X\omega^3.
\end{align}

The topological string in a generalized complex space -- topological $\CalJ$-model\footnote{$\CalJ$
stands for a generalized complex structure, which can in particular be an
ordinary symplectic (A) or an ordinary complex (B).} --- is described
in~\cite{Pestun:2005gs}, see there a complete list of references on the related works.
In agreement with~\cite{Hitchin:2} it is explained in~\cite{Pestun:2005gs},
that in the case $\dim_{\BC}X = 3$ the
moduli space of geometrical deformations of  a generalized complex
structure is a special Kahler manifold. It is also shown that the topological
string three-point function is the third derivative $C_{IJK}= \p_I \p_J \p_K \CalF$ of the
holomorphic prepotential $\CalF$ of that special geometry. The manifold
of the geometrical deformations of a generalized complex structure is
a holomorphic Lagrangian submanifold
inside the total extended moduli space of deformations of the associated
special differential BV algebra.
The outcome of~\cite{Pestun:2005gs} is that the genus zero topological
string free energy without instanton corrections is given by the
same formula $\CalF = \frac 1 2 X^I F_I$, where $X_I$ and $F^I$ are
periods the canonical pure spinor that defines a generalized complex structure
over extended set of cycles on $X$.
Therefore the relation
\begin{align}
  S_{Hit} (X',F') = \mathrm{Legendre} [\CalF''(X',X''),  F' =\p_{X''} \CalF'' ]
\end{align}
holds in the generalized complex case, and $\CalF''$ is the imaginary part of the  free energy
of the topological $\CalJ$-model~\cite{Pestun:2005gs}.

What about the black hole entropy? On the one hand, given a special Kahler
geometry,
we can formally write down an appropriate $\CalN=2$ four-dimensional supergravity, and then
the relation~\eqref{eq:HitLeg} automatically holds due to the special Kahler geometry relations.
But can it physically be related to topological strings in generalized complex space?
The answer seems to be yes, and the connection is again realized by the
ten-dimensional type II string
theory compactified on the given generalized CY manifold $X$.
Recently non Calabi-Yau compactifications
were studied in much details in\cite{Grana:2004bg,Grana:2005jc,Chuang:2005qd,Fidanza:2003zi,Lindstrom:2004iw}.
We expect that the type II ten-dimensional string theory compactified on a generalized
CY manifold $X$ is related to the topological string on $X$ exactly in the same fashion
like in the usual case.
In~\cite{Witt:2005oct,Grana:2005ny} the direct relation between Hitchin functionals for generalized complex
geometry in $\CalN=2$ supergravity and the type II string compactification was described.

\section{Examples: $T^6$ and $T^2\times K3$}
\label{sec:ex}
Here we consider a simple example when the generalized Hitchin
functional differs from the ordinary Hitchin functional at tree level.
This is possible only when $X$  has $b_1(X) \neq 0$, so $T^6$ and $T^2 \times K3$
are natural examples to see explicitly how the generalized Hitchin functional works.

First of all, one shall note that the physical type II string
compactified on $T^6$ or $T^2 \times K3$ space gives rise to $\CalN=8$
or $\CalN=4$ supergravity. Of course, the structure of these gravity theories differs from
$\CalN=2$. The usual, or generalized like in~\cite{Pestun:2005gs}
topological string, as well as attractor equations,
deals only with $\CalN=2$ terms.

The additional massless vector multiplets of $\CalN=4$ or $\CalN=8$ gravities
are not among observables of the topological string, which couples to variations
of (generalized) complex or symplectic structure on $X$.
We consider $\CalN=2$ truncation of the $\CalN=4,8$ theories and leave only those vector
multiplets, whose scalars come from (generalized) complex or symplectic moduli of $X$.

In $T^6$ case the $\CalN=8$ supergravity
multiplet~\cite{} contains the following $\CalN=2$ multiplets.
There is 1 $\CalN=2$ gravity multiplet, 6 $\CalN=2$ gravitini multiplets,
15 $\CalN=2$ vector multiplets
and 10 $\CalN=2$ hypermultiplets. Each gravitini multiplet has two gauge fields, so there are in
total $1+12+15=28$ gauge fields for the $T^6$ compactification. We throw away the gravitini multiplets
and stay with $1+15=16$ gauge fields coming from the $\CalN=2$ supergravity sector.

In $T^2 \times K3$ case, after decomposition under $\CalN=2$,
the gauge fields are counted as follows. There is one $\CalN=4$ supergravity multiplet
and 22 $\CalN=2$ vector multiplets. The $\CalN=4$ supergravity multiplet is decomposed
into 1 $\CalN=2$ gravity multiplets, 2 $\CalN=2$ gravitini multiplets and 1 $\CalN=2$ vector multiplet. It has
$1+4+1 = 6$ gauge fields. In total there are $22+6=28$ gauge fields with corresponding
28 magnetic and 28 electric charges.  Again we throw away the $\CalN=2$ gravitini multiplets
and stay with $1+22+1=24$ gauge fields coming from $\CalN=2$ supergravity sector.

The corresponding black hole solution carries magnetic and electric charges only for these $\CalN=2$ vector
multiplets. The solution is $1/4$ BPS for $T^6$ compactifications and
$1/2$ BPS for $T^2 \times K3$, so it preserves $\CalN=1$ four-dimensional supersymmetry.
The truncation is consistent classically, and we will work here only at the classical level.

Though in derivation of the Legendre transform we closely follow~\cite{Pioline:2005vi},
the novelty is the relation of the result with the generalized Hitchin functional~\cite{Hitchin:2}
and with the generalized topological $\CalJ$-model~\cite{Pestun:2005gs}.

The simplest case is the 1/8 BPS black hole for the IIA on $T^6$ with the charges corresponding
to $D0,D2,D4$ and $D6$ branes\cite{Shih:2005uc,Shih:2005qf,Shih:2005he,Pioline:2005vi,Dabholkar:2005by,Dabholkar:2005dt,Dabholkar:2004yr,Parvizi:2005aa}.
The IIA corresponds to the topological $A$-model.
The  genus zero topological string free energy is given by
\begin{align}
\label{eq:TF}
  \CalF =  - \frac 1 6 \frac {C_{IJK} {X^I X^J X^A}} {X^0},
\end{align}
where $X^I = \int_{A^I} \omega$ are integrals of the complexified Kahler class over two cycles $I=1\dots 15$,
and $C_{IJK}$ is the intersection matrix for the two-cycles on $T^6$. We will consider $A_I$ cycles
to be the $0$-cycle and all $2$-cycles, the dual $B^I$-cycles are all $4$-cycles and the $6$-cycle.
The $2$-cycles on $T^6$ are labelled by pairs $1 \leq i < j \leq 6$, which we can organize into
labels of the components of $6 \times 6$ antisymmetric matrix.
The periods $X^I$, $I=1\dots 15$  are entries of this matrix. The non-zero intersection of three 2-cycles
correspond to the choice of three pairs of indexes $(i,j)$ such that all of them are different,
with an appropriate sign coming from parity
of permutation $(i_1,i_2), (i_3,i_4), (i_5,i_6)$ into $(1,2,3,4,5,6)$.
Therefore the expression~\eqref{eq:TF} can be reorganized into the Pfaffian of the
antisymmetric $6\times 6$ matrix $X$ with entries $X^I$
\begin{align}
\label{eq:TF2}
  \CalF = - \frac {\Pf(X)} {X^0}.
\end{align}

Now we need to find the Legendre transform of~\eqref{eq:TF2} in imaginary part $X^I{}''$ for
$X^I = X^I{}' + i X^I{}''$, $I=0\dots 15$.
In order to do that for a general cubic prepotential of $n$ variables
one has to solve a system of $n$ quadratic equations, which generally speaking
does not have a closed algebraic solution~\cite{Shmakova:1996nz}. The key property of~\eqref{eq:TF2} that allows
to explicitly find its Legendre transform in $( X^0{}'', X^I{}'')$ is its extremely simple behavior
under the full complex Legendre transform for all variables $(X^I, X^0)$ at once.
There exist a very distinguished set of cubic prepotentials $ C_{IJK} X^I X^J X^K /X^0$ that
are invariant under the Legendre transform in all variables.
They were all algebraically classified in~\cite{Etingof:0003009} with even stronger condition.
The exponents of these functions are invariant under
the Fourier transform. In the $T^2 \times K3$ case, the invariance is easy to see, and we will
demonstrate it below. As for the $T^6$ case, see~\cite{Pioline:2005vi,Pioline:2003uk,Etingof:0003009}.
(The semiclassical evaluation of the Fourier transform
reduces to the Legendre transform. In other words,
integrals of exponents of such cubic functions are
exactly localized on the their critical points.\footnote{The same distinguished types of
cubic prepotentials were also classified much earlier in~\cite{Gunaydin:1983bi} studies
of $\CalN=2$ supergravity.They can appear as $\CalN=2$ four dimensional prepotentials of dimensional
reduction $\CalN=2$ five-dimensional supergravity.} Such nice prepotentials are labelled
by $B_n,D_n, E_6, E_7, E_8, F_4, G_2$ algebraic types~\cite{Pioline:2005vi}, and the case with Pfaffian
of an antisymmetric $6\times 6$ matrix is the $E_7$ case.)

Given such an invariant function $I_3(X^I)/X^0$, Pioline~\cite{Pioline:2005vi} computes
its Legendre transform in $(X^{0}{}'',X^{I}{}'')$.
The idea is to shift variables $x^I = X^{I}{}'' - \frac{X^0{}''} {X^0{}'} X^I{}'$ in
 such a way to kill the quadratic term in $X^{I}{}''$
in the expansion $I_3(X^I{}'+iX^I{}'')$. Then the Legendre transform is computed using the invariance
of $I_3(X^{I}{}'') / X^0{}''$.
In the notations $p=X',q=F',\phi = X''$ the Pioline result~\cite{Pioline:2005vi} is
\begin{multline}
S_{Hit} = \mathrm{Legendre}[-\I \frac {I_3(p^I+i\phi^I)}{p^0+i\phi^0},\phi^I] = \\
 =\sqrt{ 4p^0 I_3(q) - 4 q_0 I_3(p) + 4 \p^I I_3(q) \p_I I_3(p) - (p^0 q_0 + p^A q_A)^2} =: \sqrt{I_4(p,q)}.
\label{eq:HitTr}
\end{multline}

Specializing to the $T^6$ case,
Pioline~\cite{Pioline:2005vi} obtains quartic $SO(6,6)$ invariant functional
$I_4(p^I,q_I)$ of 32 charges $p^I,q_I, I=0\dots 16$.
The charge vector $p^I, q_I$ of $T^6$ transforms as a spinor under $SO(6,6)$, and $I_4(p_I,q^I)$
is the singlet in the symmetric tensor product of four $SO(6,6)$ spinors.

Now recall the definition~\eqref{eq:HitGen} of the generalized Hitchin functional~\cite{Hitchin:2}.
Specializing to the case of $T^6$, where in the critical point $\Omega = \rho + i \hat \rho$ is constant,
one immediately recognizes the agreement with the Legendre transform~\eqref{eq:HitTr} of the
topological string free energy~\eqref{eq:TF2}. In the framework of the generalized topological
strings~\cite{Pestun:2005gs}, the
periods $(X^I, F_I)$ are defined by integrals of the canonical pure spinor $\Omega=\rho + i \hat \rho$ of
$SO(TX,T^*X)$, equivalently it is a  mixed differential form on $X$. After the Legendre
transform the charges $(p^I,q_I)$ are identified with the periods of the real part $\rho$ of $\Omega$.
In the case of $A$-model, $\Omega = e^{i \omega + b}$, which gives the claimed correspondence.

What about the generalized $B$-model on $T^6$? An ordinary complex structure is defined  by
a holomorphic $(3,0)$ form. A generalized complex structure is defined by
a pure $SO(6,6)$ spinor of odd chirality, which can be represented as a mixed differential
 form\footnote{In this correspondence gamma matrices of $SO(TX,T^*X)$ are organized into
creation and annihilation operators $a^{i+},a_j$,  $\{a^i{}^+,a_j\} =\delta^{i}_j$.
Then $a^i{}^+ \simeq dx^{i}\wedge$ corresponds to the wedge product with $dx^i$,
and $a_i{} \simeq \p_i$ corresponds to the contraction with the vector field $\p_i$.}  $\Omega = \Omega^{(1)} + \Omega^{(3)} + \Omega^{(5)}$. The condition `pure'
for the $SO(6,6)$ spinor $\Omega$ in the generalized complex case is an analogue
of the $(3,0)$ type condition for the form $\Omega$ in the ordinary complex case.
Deformations of an ordinary complex structure are parameterized
by Beltrami differentials $\mu_{\bar j}^i$, so that $\p_{\bar j} \to \p_{\bar j} + \mu_{\bar j}^i \p_i$ and
$\Omega \to e^{-\mu} \Omega$. In the generalized complex case deformations are given\footnote{
At an arbitrary reference point the geometrical
deformations  in the topological $\CalJ$-model
are given by $\Lambda^{2}(L^*)$, and all  extended deformations  are given
by $\Lambda^{\bullet}(L^*)$, where $L$ is the $+i$-eigenbundle of the generalized complex structure
$\CalJ \in \End(TX\oplus T^*X)$.}
by  $\mu^{ij} + \mu^i_{\bar j} + \mu_{\bar i \bar j}$, which can be viewed as a section of
$\Lambda^2(TX^{10} \oplus T^*X^{01}) =: \Lambda^{2}(L^*)$, that is a subalgebra of $so(6,6,\BC)$.
A deformation $\Omega \to e^{-\mu} \Omega$ is a rotation of a spinor by an
element $\mu$ of $\Lambda^{2}(L^*) \subset so(6,6,\BC)$. We restrict $so(6,6,\BC)$ to $\Lambda^{2}(L^*)$
to keep the spinor pure.
Let us introduce indexes $(a,b)$ which run over upper holomorphic ${}^{123}$ indexes and
lower antiholomorphic ${}_{\bar 1 \bar 2 \bar 3}$ indexes.
Then an element $\mu_{ab}$ of $ \Lambda^{2}(L^*) \subset so(6,6,\BC)$ defines a rotation of the spinor $\Omega$ by the formula
\begin{align}
  \Omega = e^{-\mu} \Omega_0 = e^{-\frac 1 2 \mu_{ab} \Gamma^{a} \Gamma^{b}} \Omega_{0}.
\end{align}

The entries $\mu_{ab} = (\mu^{ij}, \mu^{i}_{\bar j}, \mu_{\bar i \bar j})$ are organized
into  an antisymmetric $6\times 6$ matrix
\begin{align}
\mu_{ab} = \left( \begin{array}{cc}
  \mu^{ij} & \mu^i_{\bar j} \\
  -(\mu_{\bar j}^i)^{T} & \mu_{\bar i \bar j}
 \end{array} \right ).
\end{align}

In the case of $T^6$  deformations, $\mu$ is a constant matrix,
and the general Chern-Simons like cubic formula~\cite{Pestun:2005gs} for the tree level  free energy of $\CalJ$-model
is reduced to
\begin{align}
 \CalF(\mu|\Omega_0) = -\frac 1 6 ((\mu_{ab} \Gamma^a \Gamma^b)^{3} \Omega_0, \Omega_0),
\end{align}
which in turn gives
\begin{align}
 \CalF{(\mu|\Omega_0)} = -\Pf(\mu).
\end{align}

We see that in the canonical coordinates, the free energy of the $B$-model on $T^6$ is also given
by the cubic polynomial, namely Pfaffian of an antisymmetric $6\times 6$ matrix. We can
also write the formula in terms of periods $X^I = \int_{A_I}( \mu \cdot \Omega)$,
where 15 $A_I$ cycles in $(H_{1} \oplus H_{3} \oplus H_{5})(X,\BC)$ are dual to the forms $\mu \cdot \Omega$ as follows.
There are 3 one-cycles for $dz^i$, 9 three-cycles $dz^{\bar i}
\wedge dz^j \wedge dz^k$,
and 3 five-cycles $dz^1\wedge dz^2\wedge dz^3 \wedge dz^{\bar i} \wedge dz^{\bar j}$.
In addition there is one distinguished cycle $A_0$, which is dual to $dz^1 \wedge dz^2 \wedge dz^3 $.
In terms of these periods $X_I = \int_{A^I} \ (\mu \cdot \Omega)$ we obtain
\begin{align}
 \CalF= -\frac{ \Pf(X)} {X^0}.
\end{align}
Then one proceeds in a similar way to the $A$-model considered above.
For an illustration let us look at the Hodge diamond of $T^6$.
The spaces of $\Omega$ and  $(\mu \cdot \Omega)$, which describe
 deformations of generalized complex structure with a reference point being
an ordinary complex structure, are underlined. They are mirror to $H^0 \oplus H^2$ in the $A$-model
by 90 degree rotation of the Hodge diamond
\begin{align}
\begin{array}{ccccccc}
         & & & h^{00} & & &  \\
       & & h^{10} &  & h^{01}  & & \\
    & h^{20} &  &  h^{11}   & & h^{02} & \\
    h^{30}  & &  h^{21}  & & h^{12} & & h^{03} \\
    & h^{31} & & h^{22}  & &   h^{13}   & \\
      & & h^{32}  & & h^{23} & & \\
       & & &  h^{33} & & & \\
\end{array} =
\begin{array}{ccccccc}
         & & & 1\,\, & & &  \\
       & & \underline{3}\,\, &  &  3\,\, & & \\
    & 3\,\, &  &  9\,\,   & & 3\,\, & \\
    \underline{1}\,\,  & & \underline{9}\,\,  & &  9\,\, & & 1\,\, \\
    & 3\,\, & & 9\,\,  & &   3\,\,   & \\
      & & \underline{3}\,\,  & &  3\,\, & & \\
       & & &  1 & & & \\
\end{array}.
\end{align}

Let us remark however, that such a simple cubic formula for $\CalF$ of
the generalized $B$-model
is obtained only in the so called canonical coordinates $\mu$,
in terms of periods $X^I$ over carefully chosen set of cycles by the condition that
$X^I$ are linear functions of $\mu$. And the fact that $\CalF$ of the $B$-model on $T^6$ does not
have corrections to the cubic term by mirror symmetry means the well-known fact
that the topological $A$-model on $T^6$
does not have instanton contributions, so the formula~\eqref{eq:TF2} is exact in genus zero\footnote{Actually,
the higher genus contributions also vanish.}.
The function $\CalF = \frac 1 2 X^I F_I $ is not $Sp(2N)$ invariant under
a change of basis of cycles, but $(X^I,F_I \equiv \p_I \CalF)$ transforms as a fundamental of $Sp(2N)$.
One can also compare the present computation with computation
of ordinary deformations of complex structure parameterized by $H^{2,1}(T^6)$
in~\cite{Moore:1998pn}.

Let us turn to the type II string on $K3\times T^2$~\cite{Cvetic:1995bj,Cvetic:1995uj,Shih:2005uc,Shih:2005he,Shih:2005qf,Dabholkar:2005by,Dabholkar:2004yr,Dabholkar:2005dt,Parvizi:2005aa,Pioline:2005vi}.
As we explained above,
we consider the truncation of the spectrum to $1+23=24$ gauge fields with $24$
electric and $24$ magnetic charges. The gauge fields come from reduction of RR $(p+1)$-forms  on $p$-cycles on $X$. In type IIA
$p$ is even, and in type IIB $p$ is odd.
The Hodge diamond for $\dim H^{p,q}$ of $T^2 \times K3$ has the following
form
\begin{align}
\begin{array}{ccccccc}
         & & & h^{00} & & &  \\
       & & h^{10} &  & h^{01}  & & \\
    & h^{20} &  &  h^{11}   & & h^{02} & \\
    h^{30}  & &  h^{21}  & & h^{12} & & h^{03} \\
    & h^{31} & & h^{22}  & &   h^{13}   & \\
      & & h^{32}  & & h^{23} & & \\
       & & &  h^{33} & & & \\
\end{array} =
\begin{array}{ccccccc}
         & & & 1 & & &  \\
       & & \underline{1} &  &  1 & & \\
    & 1 &  &  21  &  & 1 & \\
  \underline{ 1}  & & \underline{21} & & 21 & & 1 \\
    & 1 & & 21 & & 1 & \\
      & &\underline{1}  & &  1 & & \\
       & & &  1 & & & \\
\end{array}.
\end{align}

Again we underlined spaces of generalized deformations with a reference point being
an ordinary complex structure (the $B$-model). There are 22 ordinary CY moduli (20+1 for complex
structures on $K3$ and $T^2$, and 1 for an overall dilatation of the holomorphic $(3,0)$-form)
and 2 generalized extra moduli coming from deformations by a holomorphic bivector $\beta^{ij}$
and $B$-field $B_{\bar i \bar \j}$. After contraction with the holomorphic
$(3,0)$ form, the $\beta^{ij}$ and $B_{\bar i\bar j}$ generalized deformations
sit in $\Omega^{10}$ and $\Omega^{32}$ entries of the Hodge diamond.
We can decompose this deformation over the following basis in $H^{odd} = H^{1} + H^{3} + H^{5}$.

There is 1 deformation of complex structure on $T^2$, which after contraction with the holomorphic $(1,0)$ form on
$T^2$ is mapped to the $(0,1)$ form on $T^2$ times the holomorphic $(2,0)$ form on $K3$.
The coefficient at the form $(\overline {dz})_{T^2} \wedge \Omega_{K3}$ is called $X^1$.

There are 20 deformations of complex structure on $K3$ $\mu_{\bar j}^i$, which after contraction with
the holomorphic $(2,0)$ form are mapped into $(1,1)$ forms $(\mu \cdot \Omega_{K3})_{i \bar j}$ on
$K3$ times the holomorphic $(1,0)$ form on $T^2$. The corresponding coefficients
are called $X^I$, $I=2..21$.

There is 1 generalized deformation by holomorphic bivector on $K3$, which  after contraction with the $(3,0)$
holomorphic form $\Omega$ is mapped to the $(1,0)$ holomorphic form on $T^{2}$.
The corresponding coefficient is  $X^{22}$.

There is 1 generalized deformation by $B_{\bar i \bar j}$ field, which is mapped to the
space spanned by $(\overline{dz})_{T^2} \wedge \Omega_{K3} \wedge \bar \Omega_{K3}$.
The corresponding coefficient is $X^{23}$.

There is 1 dilatation of $\Omega$, which is mapped to the same space of $(3,0)$ forms $\Omega$.
The corresponding coefficient is $X^0$.

In total we have $24$ periods $X^I$, $I=0,\dots 23$
 corresponding to the $24$ gauge fields in $\CalN=2$ vector multiplets.

Using formalism of~\cite{Pestun:2005gs} it is not difficult to see
the topological string free energy is given by
\begin{align}
\label{eq:FK3}
 \CalF =  \frac 1 2 \frac  {X^1 C_{ab} X^a X^b}{X^0}
\end{align}
where $C_{ab}$ is the intersection matrix in $H^{2}(K3)$, $a=2\dots 23$. Again
the $B$-model answer is a simple cubic expression\footnote{
The solution of the Kodaira-Spencer equation for $\bar \p  (a+x) + \frac 1 2 \{ (a+x),(a+x)\}$
for $\mu = x+a$, gives a nonzero correction $a$ to the harmonic
representative $x$ of cohomology class $H^1(TX)$.
However the correlation $\int_{T^2\times K3} ((\mu \cdot)^3 \Omega, \Omega)$ decouples into
$\int_{T^2}$ and $\int_{K3} ((\mu \cdot)^2 \Omega_{K3}, \Omega_{K3}) = \int_{K3} (\mu \cdot \Omega) \wedge (\mu \cdot \Omega)$. That differs from $ (x\cdot \Omega) \wedge (x \cdot \Omega)$ by an integral
of $\p$-exact times $\p$-closed term, which vanishes.},
exactly in agreement with the mirror symmetry ($T^2 \times K3$ is mirror symmetric to itself)
and the fact that $A$-model does not have
any worldsheet instanton corrections in genus zero.

The full Legendre transform of the function
$\CalF = \frac 1 2 X^1 C_{ab} X^{a} X^{b}/X^{0}$ in all complex variables $X^I$
is given simply by $-\frac 1 2 F_1 C^{ab} F_a F_b/F_{0}$. (In other words, for the bilinear form
that satisfies $C=C^{-1}$, the function~\eqref{eq:FK3} is invariant under the full Legendre transform
and fall into the classification of~\cite{Etingof:0003009}).
Explicitly, we need to solve $\p_0 \CalF = F_0, \p_1 \CalF = F_1$ and $\p_a \CalF = F_a$, so we have
\begin{align}
  -\frac 1 2 \frac {X^1} {(X^0)^2} C_{ab} X^{a} X^{b} = F_0 \\
 \frac 1 2  \frac 1 {X^0} {C_{ab} X^{a} X^{b}} = F_1\\
  \frac {X^1} {X^0} C_{ab} X^{b} = F_a.
\end{align}
Dividing the first line over the second, we have $\frac {X^{0}}{X^{1}} = -\frac {F_1} {F_0}$.
From the third line we have $X^{b} = \frac {X^0}{X^1} C^{ba} F_{a} = -\frac{F_1} {F_0} C^{ba} F_{a}$.
Then $C_{ab} X^{a} X^{b} = (\frac{F_1} {F_0})^2 C^{ba}F_a F_b$. Then from the second line
we have $X^{0} = \frac 1 2 \frac {F_1} {(F_0)^2} C^{ba}F_{a} F_{b}$, and,
using $F_0/F_1 = -X^{0}/X^{1}$ we get $X^{1} = -\frac 1 2 \frac 1 {F^0} {C^{ba}F_a F_b}$.
We see that $X^I$ are expressed in terms of $F_I$ in the same way as $F_I$ in terms of $X^I$
(up to the minus sign).

For any homogeneous function of weight two $\CalF = \frac 1 2 F_I X^{I}$, the Legendre
transform is given by $\tilde \CalF = F_I X^I - \CalF = \frac 1 2 F_I X^I(F_J)$.
We plug the expressions for $X^{I}$ and obtain
\begin{align}
  \tilde \CalF = -\frac 1 2 \frac {F_1 C^{ab} F_a F_b} {F_{0}}.
\end{align}

Then we can use Pioline~\cite{Pioline:2005vi} formula \eqref{eq:HitTr} for
the Legendre transform in imaginary part of $X$ to find
\begin{align}
 S_{BH} = \pi S_{Hit} = \pi \sqrt{ p^2 q^2 - (p \cdot q)^2  },
\end{align}
where the charge vectors $(p^I,q_I)$ are identified with real part of $(X^I,F_I)$,
and the scalar product is taken in the $(20,4)$ signature lattice.
This is the truncation of the full $(22,6)$ lattice for type II on $T^{2} \times K3$
to the charges of $\CalN=2$ vector multiplets
in agreement with~\cite{Cvetic:1995bj,Cvetic:1995uj,Shih:2005uc,Shih:2005he,Dabholkar:2005by,Dabholkar:2004yr,Dabholkar:2005dt,Parvizi:2005aa,Pioline:2005vi}.

\section{Conclusion}\label{sec:con}
In this note it was argued that the OSV conjecture~\cite{Ooguri:2004zv} is applicable to the case
of generalized complex structures~\cite{Hitchin:2,GCS}. If $b_1(X)=0$ one has to use
generalized Hitchin functional at quantum level~\cite{Pestun:2005rp}; classically
the generalized and the ordinary geometry does not differ. However, if $b_1(X) \neq 0$,
like in the case of $T^6$ or $T^2 \times K3$, the emergence of the generalized Hitchin functional
is inevitable at tree level.

Deformations of a generalized complex structure on a three-fold $X$ are parameterized by the half
of all even/odd cycles in type $A/B$.
The extra moduli exist at the classical level if $H^1(X)$ is not trivial.
For example, the extra deformations of complex structure include $H^{1,0}$ and $H^{3,2}$ in
addition to the standard $H^{2,1}$. The classical black hole entropy in this case
is given by the \emph{generalized} Hitchin functional of the form $\int \sqrt{I_{4}(\rho)}$,
where $\rho$ is the real part of the canonical pure $SO(6,6)$ spinor (mixed differential form) on $X$.
The scalar fields in $\CalN=2$ multiplets come from all such generalized deformations,
and the corresponding gauge fields come from reduction of all odd/even RR forms $C_{p}$ in type II $A/B$
on all even/odd cycles in $X$.

We did not touch  extremely interesting subjects of higher genus and nonperturbative
corrections to the relation, but suggest that the generalized geometry must be an appropriate
framework for study of the subject.
Especially this is interesting in the context
of non CY background compactifications~\cite{Grana:2005ny,Grana:2005jc,Grana:2004bg,Chuang:2005qd,Lindstrom:2004iw,Fidanza:2003zi}. The microscopical counting of black hole entropy  could illuminate
non-perturbative structure of the topological $\CalJ$-model~\cite{Pestun:2005gs}.

\acknowledgments
I would like to especially thank E. Witten for very useful comments and
suggestions. I also learned much from communication with N.~Nekrasov, to whom I am very grateful.
I thank M.~Grana, F.~Denef, A.~Kapustin, A.~Losev, A.~Neitzke, S.~Shatashvili and  D.~Shih for interesting discussions.
Part of this research was done during my visits
to Institut des Hautes Etudes Scientifiques, Bures-sur-Yvette, France and
the 3rd Simons Workshop in Mathematics and Physics at Stony Brook University, NY.
I thank these institutions for their kind hospitality.
The work was supported in part by grant RFBR 04-02-16880 and grant NSF 245-6530.

\bibliographystyle{myunsrt}

\bibliography{bsample}

\end{document}